\documentclass[fleqn,twoside]{article}
\usepackage{espcrc2}

\newcommand{\ep}{\varepsilon}
\newcommand{\nn}{\nonumber}
\newcommand{\tfrac}[2]{{\textstyle{\frac{#1}{#2}}}}
\newcommand{\Li}[2]{{\mbox{Li}}_{#1}\left(#2\right)}

\title{
\vspace*{-40mm}
\begin{flushright}
{\large MZ-TH/01-41}\\
{\large hep-ph/0210171}\\
{\large September 2002}
\end{flushright}
\vspace*{15mm}
Analytical evaluation of certain on-shell two-loop 
       three-point diagrams%
\thanks{Partially supported by the DFG, as well as by
INTAS grant 00--00313 and RFBR project 01--02--1617.}}

\author{A. I. Davydychev\address[Mainz]{Department of Physics, 
University of Mainz, Staudingerweg 7, D-55099 Mainz, Germany}%
\address[Moscow]{Institute for Nuclear Physics, Moscow State University,
119992 Moscow, Russia}%
\thanks{Current address: 
Schlumberger, 
SPC, 110 Schlumberger Dr., MD-5, Sugar Land, TX~77479, USA}
        \ and \
        V. A. Smirnov\addressmark[Moscow] }
       
\begin{document}

\begin{abstract}
An analytical approach is applied to the calculation of
some dimensionally-regulated two-loop vertex diagrams 
with essential on-shell singularities. 
Such diagrams are important for the evaluation
of QED corrections to the muon decay, QCD corrections to
top quark decays $t\to W^{+}b$, $t\to H^{+}b$, etc.
\vspace{1pc}
\end{abstract}

\maketitle

\section{INTRODUCTION}

We consider the two-loop diagram shown in Fig.~1, using  
the one-loop case as an example.
All external momenta are ingoing, $P+p+q=0$, and satisfy 
$P^2=M^2$, $p^2=m^2$. We denote 
\begin{eqnarray}
J&\!\!\equiv\!\!& 
\int{{\rm d}^n k}
\left\{k^2 \left[k^2\!+\!2(Pk)\right] 
\left[k^2\!-\!2(pk)\right]\right\}^{-1} \!\! ,
\\
F &\!\!\equiv\!\!& \int \!\! \int {\rm d}^n k \; {\rm d}^n l
\left\{ \left[ k^2\!+\!2(Pk)\right] \left[ l^2\!+\!2 (Pl)\right]
\right\}^{-1}
\nonumber \\ &&
\times \!
\left\{\left[ k^2\!-\!2(pk)\right] 
\left[ l^2\!-\!2 (pl)\right] k^2 (k\!-\!l)^2 \right\}^{-1} \!\! ,
\label{def_F}
\end{eqnarray}
where $n=4-2\ep$ is the space-time dimension.

We assume that $m^2 \ll M^2, |q^2|$ and expand the vertices
in the ratio of $m$ and the large parameters. We apply
the so-called strategy of expansion by regions~\cite{BS,SR,S1,Sb},
confining ourselves to the leading power term of the expansion
(including all logarithms and the constant part).

\newcommand{\triangleMm}
 {\setlength {\unitlength}{0.7mm}
 \begin{picture}(36,20)(0,28)
\thicklines
 \put (18,48) {\line(0,1){6}}
 \put (18,48) {\line(-1,-3){12}}
\thinlines
 \put (18,48) {\line(1,-3){12}}
 \multiput(8.5,18)(2,0){10}{\line(1,0){1}}
 \put (18,48) {\circle*{1}}   
 \put (8,18)  {\circle*{1}}
 \put (28,18) {\circle*{1}}
 \put (6,30)  {\makebox(0,0)[bl]{\large $M$}}
 \put (26,30) {\makebox(0,0)[bl]{\large $m$}}
 \put (17,13)  {\makebox(0,0)[bl]{\large $0$}}
 \put (0,12) {\makebox(0,0)[bl]{\large $P$}}
 \put (32,11) {\makebox(0,0)[bl]{\large $p$}}
 \put (14,50) {\makebox(0,0)[bl]{\large $q$}}   
 \end{picture}}
 
\newcommand{\vertexMm}
 {\setlength {\unitlength}{0.7mm}
 \begin{picture}(36,20)(0,28)
\thicklines
 \put (18,48) {\line(0,1){6}}
 \put (18,48) {\line(-1,-3){12}}
\thinlines
 \put (18,48) {\line(1,-3){12}}
 \multiput(13.5,33)(2,0){5}{\line(1,0){1}}
 \multiput(8.5,18)(2,0){10}{\line(1,0){1}}
 \put (18,48) {\circle*{1}}   
 \put (13,33)  {\circle*{1}}
 \put (23,33) {\circle*{1}}
 \put (8,18)  {\circle*{1}}
 \put (28,18) {\circle*{1}}
 \put (4,25)  {\makebox(0,0)[bl]{\large $M$}}
 \put (27,25) {\makebox(0,0)[bl]{\large $m$}}
 \put (9,40)  {\makebox(0,0)[bl]{\large $M$}}
 \put (22,40) {\makebox(0,0)[bl]{\large $m$}}
 \put (17,13)  {\makebox(0,0)[bl]{\large $0$}}
 \put (17,28) {\makebox(0,0)[bl]{\large $0$}}
 \put (0,12) {\makebox(0,0)[bl]{\large $P$}}
 \put (32,11) {\makebox(0,0)[bl]{\large $p$}}
 \put (14,50) {\makebox(0,0)[bl]{\large $q$}}   
 \end{picture}}

\begin{figure}[b]
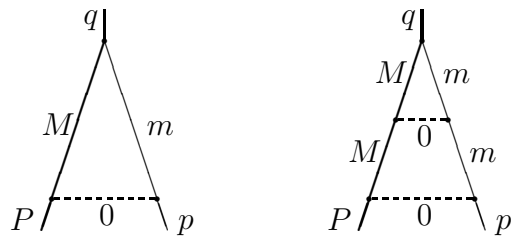

\[
\begin{array}{ccc}
\triangleMm & $\hspace{1cm}$ & \vertexMm
\end{array}
\]
\caption{One- and two-loop vertex diagrams}
\end{figure}

In section~2 of~\cite{S1} (the second example), 
this approach was applied to $F\big|_{q^2=0}$.
Introduce $n_{1,2}=(\tfrac{1}{2},\mp \tfrac{1}{2}, \underline{0})$,
so that $2(n_{1,2}k)\equiv k_{\pm} = k_0\pm k_1$.
Here ``underlined" means all remaining components: e.g., 
$k=(k_0,k_1,\underline{k})$.
It is convenient to choose $P=(-M,0,\underline{0})$ and
$p=\alpha n_1 + (m^2/\alpha)n_2$, with
$
2M\alpha= M^2+m^2-q^2
+[\lambda(M^2,m^2,q^2)]^{1/2},
$
where $\lambda(M^2,m^2,q^2)=[(M+m)^2-q^2][(M-m)^2-q^2]$. 
Then, the  relevant regions are
\begin{eqnarray}
\mbox{{\em hard} (h):}
&& \hspace*{-5.5mm} k \sim M\,;
\nn 
 \\
\mbox{{\em 1-collinear} (1c):}
&& \hspace*{-5.5mm}
k_+ \sim m^2/M ,\,\,k_-\sim M\, , \,\, {\underline{k}} \sim m\,;
\nn 
 \\
\mbox{{\em 2-collinear} (2c):}
&& \hspace*{-5.5mm}
k_+\sim M,\,\,k_-\sim m^2/M\, ,
\,\,{\underline{k}} \sim m \,;
\nn 
\\
\mbox{{\em ultrasoft} (us):}
&& \hspace*{-5.5mm}
k \sim m^2/M\, .
\nn 
\end{eqnarray}
Note the change ${\rm (1c)}\leftrightarrow{\rm (2c)}$,
as compared to~\cite{S1}.

For the one-loop diagram $J$, only the (h) and (1c) contributions
are relevant in the leading order of the expansion in $m$.
For the two-loop diagram $F$, the following ($k$-$l$) regions 
yield non-zero contributions: 
(h-h), (1c-h), (1c-1c) and (us-1c). 

\section{ONE-LOOP DIAGRAM}

In the on-shell case, the three-point function $J$
reduces to a two-point function with 
the space-time dimension $2-2\ep$,
multiplied by $\pi/(2\ep)$ (see section~3.2 of~\cite{DK1}).
Therefore, the two-point function should be expanded in $\ep$ 
up to the next term. 
In fact, any term of the $\ep$ expansion can be calculated
in terms of the log-sine functions (see in~\cite{D-ep,DK1})
whose analytic continuation yields Nielsen polylogarithms $S_{a,b}(z)$
(see in~\cite{DK-Bastei}).

Expanding exact result in $\ep$ and $m^2$, we obtain
\begin{eqnarray}
\label{expanded2}
J &\!\!=\!\!&
{\rm i} \pi^{2-\ep} 
{\rm e}^{-\gamma_{\rm E} \ep}
(M^2)^{-1-\ep} \sigma^{1+2\ep}
\bigl\{ -\tfrac{1}{2}\ep^{-1} L
\nonumber \\ &&
+\tfrac{1}{4}L^2
-\Li{2}{u} \bigl\} 
+{\cal{O}}(\ep, m^2L^2) ,
\end{eqnarray}
where $\sigma\equiv M^2/(M^2-q^2)$,
$u\equiv 1-1/\sigma=q^2/M^2$,
$L\equiv \ln\left(m^2\sigma^2/M^2\right)$.
Let us check whether the sum of the (h) and (1c) contributions
gives the same. For the (h) contribution, $J\big|_{m=0}$, we get
\begin{eqnarray}
\label{int16}
J\big|_{m=0}
&\!\!\!=\!\!& 
{\rm i}\pi^{2-\ep} \Gamma(1+\ep) (M^2)^{-1-\ep} \sigma^{1+2\ep}
\nonumber \\ && 
\times\!
\bigl\{ -\tfrac{1}{2}\ep^{-2} + \ln^2{\sigma}
+\Li{2}{u} \bigl\}
+{\cal{O}}(\ep) .
\end{eqnarray}
Adding to (\ref{int16}) the (1c) contribution (see in~\cite{S1}),
$
{\mbox{i}}\pi^{2-\ep}\Gamma(1+\ep)\;(M^2)^{-1-\ep}\;
(m^2)^{-\ep}\; \sigma/(2\ep^2)
$,
and expanding in $\ep$, we reproduce Eq.~(\ref{expanded2}).


\section{TWO-LOOP DIAGRAM WITH $m=0$}

The (h-h) region generates Taylor expansion in $m^2$ 
of the integrand of $F$, yielding $F\big|_{m=0}$ 
in the leading order.  
Here, we could not employ the methods of~\cite{other1,other2}
which were useful when $P^2=p^2=0$. Instead,
using~\cite{BD-TMF} we derived
a four-fold Mellin--Barnes representation for $F\big|_{m=0}$,
\begin{eqnarray}
\label{MB0}
&& \hspace*{-5mm}
-\frac{\pi^{4-2\ep} (M^2)^{-2-2\ep}}{\Gamma(1-2\ep)}\;
\frac{1}{(2\pi{\rm i})^4}
\int\hspace*{-4mm}
\int\limits_{\;\;\; -{\rm i}\infty}^{\;\;\; {\rm i}\infty}
\hspace*{-4mm}\int\!\!\int
{\rm d}z\; {\rm d}\tilde{z}\; {\rm d}t\; {\rm d}w\;
\nonumber \\ && \hspace*{-5mm}
\times
{\sigma}^{2+2\ep+z+\tilde{z}} 
\frac{\Gamma(-t)\; \Gamma(-w)\; \Gamma(1+t+w)}
{\Gamma(1-t)\; \Gamma(1-w)\; \Gamma(1\!-\!2\ep\!+\!t\!+\!w)} 
\nonumber \\ && \hspace*{-5mm}
\times
\Gamma(1+\ep+t+w+z)\; \Gamma(1+\ep-t-w+\tilde{z})\;
\nonumber \\ && \hspace*{-5mm}
\times
\Gamma(-z)\; \Gamma(-\tilde{z})\;
\Gamma(-\ep-t-z)\; \Gamma(-\ep+t-\tilde{z}) \;
\nonumber \\ && \hspace*{-5mm}
\times
\Gamma(-\ep-w+z)\; \Gamma(-\ep+w+\tilde{z})\; .
\end{eqnarray}
The contour integrals separate the right and left 
series of poles of $\Gamma$ functions in 
$z$, $\tilde{z}$, $t$ and $w$.
For small negative $\ep$, this can be satisfied 
by straight contours
(parallel to the imaginary axes), if we take, say,
$\mbox{Re}{z}=\mbox{Re}{\tilde{z}}=\frac{1}{2}\ep$, 
$\mbox{Re}{t}=\ep$,
$\mbox{Re}{w}=\frac{1}{4}\ep$.
In fact, one can integrate by parts~\cite{ibp} 
to shrink a line, but
this does not simplify the calculation.

The result of a tedious calculation of $F\big|_{m=0}$ is
\begin{eqnarray}
\label{resultM0}
&& \hspace*{-5mm}
\pi^{4-2\ep} {\rm e}^{-2\gamma_{\rm E}\ep} (M^2)^{-2-2\ep}
{\sigma}^{2+4\ep} 
\nonumber \\ &&
\times
\bigl\{
\tfrac{1}{12}\ep^{-4} +
\ep^{-2} 
\left[ \tfrac{1}{12}\pi^2 
      + \tfrac{1}{2}\Li{2}{u} \right] 
\nonumber \\ &&
+ \ep^{-1} 
\left[ 
\tfrac{91}{36}\zeta_3 - S_{1,2}(u) + \tfrac{3}{2} \Li{3}{u} 
\right] 
\nonumber \\ &&
+ \tfrac{179}{1440}\pi^4 
+ \tfrac{7}{12}\pi^2 \Li{2}{u}
- \big[\Li{2}{u}\big]^2 
\nonumber \\ &&
+ 2 S_{1,3}(u)
+ S_{2,2}(u)
+ \tfrac{5}{2} \Li{4}{u} \bigl\}
+{\cal O}(\ep) .
\end{eqnarray}


\section{TWO-LOOP DIAGRAM WITH $m\neq 0$}

The (1c-1c) and (us-1c) contributions can be 
trivially obtained from
those for the limit with $q^2=0$ \cite{S1,Sb},
substituting $M^2\to M^2-q^2$.
To calculate the leading-order (1c-h) contribution,
one can apply the technique
of $\alpha$ parameters and the Mellin--Barnes representation.
In this way, the
problem is reduced to a two-fold contour integral 
which can be evaluated by the standard
technique of taking residues and shifting contours.

Collecting all contributions, we obtain for $F$
\begin{eqnarray}
\label{resultMm}
&& \hspace*{-5mm}
\pi^{4-2\ep} {\rm e}^{-2\gamma_{\rm E}\ep} (M^2)^{-2-2\ep}
{\sigma}^{2+4\ep} \bigl\{
\tfrac{1}{8}\ep^{-2} L^2
\nonumber \\ &&
-\ep^{-1}
\bigl[
\tfrac{1}{6}L^3
+\tfrac{1}{12} \pi^2 L
-\tfrac{1}{2} L\; \Li{2}{u}
+\zeta_3
\bigl]
\nonumber \\ &&
+\tfrac{13}{96} L^4
+\tfrac{5}{16} \pi^2 L^2
-\tfrac{1}{4} L^2\; \Li{2}{u}
\nonumber \\ &&
- L\; S_{1,2}(u)
+\tfrac{3}{2} L\; \Li{3}{u}
+\tfrac{3}{2} \zeta_3 L
-2 S_{2,2}(u)
\nonumber \\ &&
+\tfrac{1}{2} \left[ \Li{2}{u} \right]^2
+\tfrac{1}{72}\pi^4 \bigl\} 
+ {\cal O}(\ep, m^2 L^4)\; .
\end{eqnarray}
Note that the $\ep^{-4}$ and $\ep^{-3}$ terms have cancelled.
Eqs.~(\ref{resultM0})--(\ref{resultMm})
were checked numerically, using~\cite{BH}.
At $q^2\!=\!0$, they reproduce
Eqs.~(15) and (21) of~\cite{S1}.


\end{document}